\documentstyle[12pt,epsf]{article}								 
\def\CellGroup{\bgroup}
\def\endCellGroup{\egroup}
\begin{document} 
\vspace*{3cm}
\begin{center}
{\large\bf Muon transverse polarization in the $K_{l2\gamma}$ decay
in SM}
\end{center}

\vspace*{1cm}
\begin{center}
Braguta V.V.$^{\dagger}$, Chalov A.E.$^{\dagger}$, Likhoded A.A.$^{\dagger\dagger}$
\end{center}

\vspace*{1cm}
\begin{center}
$^{\dagger}$ {\it Moscow Institute of Physics and Technology, Moscow, 141700  Russia}
\\
$^{\dagger\dagger}$ {\it Instituto de F\'isica Te\'orica - UNESP, \\
 Rua Pamplona, 145, 01405-900 S\~ao Paulo, SP, Brasil}\footnote{
On leave of absence from Institute for High Energy Physics, Protvino, 142284 Russia}\\
andre@ift.unesp.br
\end{center}

\vspace*{2cm}
\underline{\bf Abstract}

\vspace*{0.5cm}
\noindent
The muon transverse polarization in the  
$K^+ \to \mu^+ \nu \gamma$ process  induced by the 
electromagnetic final state interaction is calculated in the framework 
of Standard Model. 
It is shown that one loop contributions lead to a nonvanihing
muon transverse polarization.
The value of the muon transverse polarization averaged
over the kinematical region of $E_\gamma \geq 20$~MeV is equal to $5.63 \times
10^{-4}$.

\newpage
\section{Introduction}

\noindent
The study of the radiative $K$-meson decays is extremely interesting
in searching for new physics effects beyond the Standard
Model (SM). One of the most appealing possibilities is to probe
new interactions, which could lead to $CP$-violation.
Contrary to SM, where the $CP$-violation is caused by the presence of the complex phase 
in the  CKM matrix,  the $CP$-violation in extended models (for instance, in models
with three and more Higgs doublets) can naturally arises due to the complex couplings
of new Higgs bosons to fermions [1]. Such effects can be detected
by using experimental observables, 
which are essentially sensitive to $T$-odd contributions. 
These observables, for instance, are $T$-odd correlation ($T=\frac{1}{M_K^3}
\vec p_{\gamma}\cdot [\vec p_{\pi} \times \vec p_l ]$) in the
$K^{\pm}\to\pi^0 \mu^{\pm}\nu\gamma$ decay [2] and muon transverse polarization ($P_T$)
in  $K^{\pm}\to \mu^{\pm}\nu\gamma$.
The search for new physics effects using  the $T$-odd correlation analysis
in the $K^{\pm}\to\pi^0 \mu^{\pm}\nu\gamma$  decay 
will be done in the proposed OKA experiment [3],
where an event sample of $7.0\times 10^5$ for the 
$K^+\to\pi^0 \mu^+\nu\gamma$ decay is expected to be accumulated.

At the moment the  E246 experiment at KEK [4] performs the 
analysis of the data on the $K^{\pm}\to \mu^{\pm}\nu\gamma$ process
to put bounds on the $T$-violating muon transverse polarization.
It should be noted that the expected value of new physics contribution to $P_T$
can be of the order of  $\simeq 7.0\times 10^{-3}\div 6.0\times 10^{-2}$ [5,6], 
depending on the type of model beyond SM.
Thus, when one searches for new physics contributions to $P_T$,
it is extremely important to estimate the effects coming from so called 
``fake'' polarization, which is caused by the SM electromagnetic final state
interactions and which is a natural background for the new interaction contributions.

In this paper we calculate muon transverse polarization
in the  $K^{\pm}\to \mu^{\pm}\nu\gamma$ process, induced by the
electromagnetic final state interaction in the one-loop approximation
of the minimal Quantum Electrodynamics.

In next Section we present the calculations of the muon transverse polarization
taking into account one-loop diagrams with final state interactions within the SM. 
Last Section summarizes  the results and  conclusions.

\section{Muon transverse polarization in the  $K^+ \to \mu^+ \nu \gamma$ process
in SM}

\noindent
The $K^+ \rightarrow \mu^+ \nu \gamma$ decay at the tree level of SM
is described by the diagrams shown in  Fig.~1. The diagrams in Fig. 1b and 1c
correspond to the muon and kaon bremsstrahlung, while the diagram in Fig. 1a 
corresponds to the structure radiation.
This decay amplitude can be written as follows
\begin{equation}
M=ie \frac{G_F}{\sqrt{2}}V^{*}_{us}\varepsilon^{*}_{\mu}\left(f_K m_{\mu}
\overline u(p_{\nu})(1+\gamma_{5}) \biggl( \frac{p_{K}^{\mu}}{(p_K q)}
-\frac{(p_{\mu})^{\mu}}{(p_{\mu} q)}-
\frac{\hat{q} \gamma^{\mu}}{2( p_{\mu} q)} \biggr)v(p_{\mu})-G^{\mu \nu}
l_{\nu}
\right) \; ,
\end{equation}
where 
\begin{eqnarray}
l_{\mu}=\overline u(p_{\nu}) (1+\gamma_{5}) \gamma_{\mu}
v(p_{\mu}) \; ,\nonumber \\
G^{\mu \nu}= i F_v ~\varepsilon^{ \mu \nu \alpha \beta } q_{\alpha}
(p_K)_{\beta} -
F_a ~ ( g^{\mu \nu} (p_K q)-p_K^{\mu} q^{\nu} )\;, 
\end{eqnarray}
$G_F$ is the Fermi constant, $V_{us}$ is the corresponding 
CKM matrix element, $f_K$ is the $K$-meson leptonic constant,
$p_K$, $p_\mu$, $p_\nu$, $q$ are the kaon, muon, neutrino, and photon four-momenta,
correspondingly, and $\varepsilon_{\mu}$ is the photon polarization vector. 
$F_v$ and $F_a$ are the kaon vector and axial formfactors. 
In Eq. (2) we use the following
definition of Levi-Civita tensor: $\epsilon^{0 1 2 3} = +1$.

The part of the amplitude which corresponds to the structure radiation 
and kaon bremsstrahlung and which will be used further in one-loop calculations,
has the form:
\begin{equation}
M_K=ie \frac{G_F}{\sqrt{2}}V^{*}_{us}\varepsilon^{*}_{\mu}\left(f_K m_{\mu}
\overline u(p_{\nu})(1+\gamma_{5}) \biggl( \frac{p_{K}^{\mu}}{(p_K q)} -
\frac {\gamma^\mu}  m_\mu  
  \biggr)v(p_{\mu})-G^{\mu \nu} l_{\nu}
\right)\;.
\end{equation}
The partial width of the $K^+ \to \mu^+ \nu \gamma$ decay in the  
$K$-meson rest frame can be expressed as 
\begin{equation}
d \Gamma = \frac {\sum |M|^2} { 2 m_K } (2 \pi)^4 \delta ( p_K - p_\mu -q
-p_\nu)
\frac {d^3 q} {(2 \pi)^3 2 E_q } \frac {d^3 p_\mu} {(2 \pi)^3 2 E_\mu }
\frac {d^3 p_\nu} {(2 \pi)^3 2 E_\nu }\;,
\end{equation}
where summation over muon and photon spin states is performed.

Introducing the unit vector along the muon spin direction in muon rest frame,
$\bf \vec s $, where ${\bf \vec e}_i~(i=L,\: N,\: T)$ are the unit vectors
along the longitudinal, normal and transverse components of muon polarization, 
one can write down the matrix element squared for  the transition
into the particular muon polarization state in the following form:
\begin{equation}
|M|^2=\rho_{0}[1+(P_L {\bf \vec e}_L+P_N {\bf \vec e}_N+P_T {\bf \vec
e}_T)\cdot \bf \vec s]\;,
\end{equation}
where $\rho_{0}$ is the Dalitz plot probability density averaged over polarization states.
The  unit vectors ${\bf \vec e}_i$ can be expressed in terms of the
three-momenta of final particles:
\begin{equation}
{\bf \vec e}_L=\frac{\vec p_{\mu}}{|\vec p_{\mu}|}~~~ 
{\bf \vec e}_N=\frac{\vec p_{\mu}\times(\vec q \times \vec p_{\mu})}
{|\vec p_{\mu}\times(\vec q \times \vec p_{\mu})|}~~~
{\bf \vec e}_T=\frac{\vec q \times \vec p_{\mu}}{|\vec q \times \vec
p_{\mu}|}\: . 
\end{equation}
With such definition of ${\bf \vec e}_i$ vectors, $P_T, P_L$, and $P_N$ denote
transverse, longitudinal, and normal components of the muon polarization,
correspondingly. It is convenient to use the following variables
\begin{equation}
x=\frac{2 E_\gamma}{m_K}\;,~~~y=\frac{2 E_\mu}{m_K}\;,
~~~\lambda=\frac{x+y-1-r_\mu}{x}\;,~~~r_{\mu}=\frac {m_{\mu}^2}{m_K^2}\;,
\end{equation}
where $E_\gamma$ and $E_\mu$ are the photon and muon energies in the kaon
rest frame. 

The Dalitz plot probability density, as a function of the  $x$ and $y$
variables, has the form:
\begin{eqnarray}
\rho_{0}=\frac{1}{2}e^2 G_F^2 |V_{us}|^2 &\cdot & \biggl( 
\frac {4 m_{\mu}^2 |f_K|^2 } {\lambda x^2} (1-\lambda) \Bigl (x^2+2 (1-r_{\mu})
(1-x-\frac {r_{\mu}} {\lambda} )\Bigr)	\nonumber\\
&+& m_K^6 x^2 (|F_a|^2+|F_v|^2) (y-2 \lambda y-\lambda x +2
\lambda^2)\nonumber\\
&+& 4  ~\hbox{Re} (f_K F_v^*) ~m_K^4 r_{\mu} \frac x {\lambda} (\lambda -1)
\hspace{0.3in} \nonumber\\
&+&4 ~\hbox{Re} (f_K F_a^*) ~m_K^4  r_{\mu} (-2 y+x+2 \frac {r_{\mu}} {\lambda}
-
\frac x {\lambda} +2 \lambda )\hspace{0.3in} \nonumber\\
&+& 2 ~\hbox{Re} (F_a F_v^*) ~m_K^6 x^2 (y-2 \lambda +x \lambda) \biggr)\;.
\end{eqnarray}
Calculating the muon transverse polarization $P_T$ we follow the original paper [7]
and assume that the decay amplitude  is $CP$-invariant,
and formfactors $ f_K $, $ F_v$, and $ F_a$ are real.
In this case the tree level muon polarization $P_T=0$.
When one-loop contributions are incorporated, the nonvanishing muon
transverse polarization can arise due to the interference of tree-level 
diagrams and imaginary parts of one-loop diagrams, induced by the electromagnetic 
final state interaction.

To calculate the imaginary parts of formfactors one can use the
$S $-matrix unitarity:
\begin{equation}
S^+ S=1
\end{equation}
and, using $ S=1+i T$, one gets
\begin{equation}
T_{f i}-T_{i f}^*=i \sum_n T^*_{n f} T_{n i}\;, 
\end{equation}
where $i,\: f, \: n$ indices correspond to the initial, final, and intermediate
states of the particle system.
Further, using the  $T$-invariance of the matrix element one has
\begin{eqnarray}
\mbox{Im} T_{f i}=\frac 1 2 \sum_n T^*_{n f} T_{n i}\;,  \\
T_{f i}=(2 \pi)^4 \delta ( P_f-P_i ) M_{f i}\;.
\end{eqnarray}
One-loop diagrams of SM, which contribute to the muon transverse polarization in
the $K^+ \to \mu^+ \nu \gamma$ decay, are shown in Fig. 2.
Using Eq. (3) one can write down the imaginary parts of these diagrams.
For diagrams in Figs. 2a, 2c one has
\begin{eqnarray}
~\hbox{Im}M_1=\frac{i e \alpha}{2 \pi}\frac{G_F}{\sqrt{2}}V_{us}^*
\overline u(p_{\nu})(1+\gamma_{5}) \int \frac{d^3 k_{\gamma}}{2
\omega_{\gamma}}
\frac{d^3 k_{\mu}}{2 \omega_{\mu}}\delta(k_{\gamma}+k_{\mu}-P)R_{\mu}\times
\nonumber\\
(\hat k_{\mu}-m_{\mu})\gamma^{\mu}\frac{\hat q+ \hat
p_{\mu}-m_{\mu}}{(q+p_{\mu})^2-m_{\mu}^2}
\gamma^{\delta} \varepsilon^*_{\delta} v(p_{\mu})\;. \label{im1}
\end{eqnarray}
For diagrams in Figs. 2b, 2d one has
\begin{eqnarray}
~\hbox{Im}M_2=\frac{i e \alpha}{2 \pi}\frac{G_F}{\sqrt{2}}V_{us}^*
\overline u(p_{\nu})(1+\gamma_{5}) \int \frac{d^3 k_{\gamma}}{2
\omega_{\gamma}}
\frac{d^3 k_{\mu}}{2 \omega_{\mu}}\delta(k_{\gamma}+k_{\mu}-P)R_{\mu}\times
\nonumber\\
(\hat k_{\mu}-m_{\mu})\gamma^{\delta} \varepsilon^*_{\delta} 
\frac{ \hat k_{\mu}-\hat q-m_{\mu}}{(k_{\mu}-q)^2-m_{\mu}^2}
\gamma^{\mu} v(p_{\mu}), \label{im2}
\end{eqnarray}
where
\begin{eqnarray}
R_{\mu}&=&f_K m_{\mu} \biggl (\frac {(p_K)_{\mu}} {(p_K k_{\gamma})} -
\frac {\gamma_{\mu}}{m_{\mu}} \biggr) -i F_v \varepsilon_{\mu \nu \alpha \beta} 
(k_{\gamma})^{\alpha} (p_K)^{\beta} \gamma^{\nu}\nonumber \\
& + & F_a (\gamma_{\mu} (p_K k_{\gamma})-(p_K)_{\mu} \hat k_{\gamma})\:.
\end{eqnarray}
To write down the contributions of diagrams shown in Figs. 2e, 2f,
one should substite $R_{\mu}$ by 
\begin{eqnarray}
R_{\mu}&=&f_K m_{\mu} \biggl (
\frac {\gamma_\mu}{m_\mu} - \frac{(k_\mu )_\mu}{(k_\mu k_\gamma )}
-\frac{\hat k_\gamma \gamma_\mu }{2(k_\mu k_\gamma )}\biggr) 
\end{eqnarray}
in expressions (\ref{im1}), (\ref{im2}).

Using $\chi$PT Lagrangian [8], one can derive  decay amplitudes
for the $K^+ \to \pi^0 \mu^+ \nu$ and $\pi^0 \to \gamma 
\gamma$ processes, which  contribute to the imaginary part of diagram in Fig. 2g:  
\begin{eqnarray}
\nonumber 
T(K^+ \to \pi^0 \mu^+ \nu) &=& - \frac {G_F} 2 \bar u (  p_{\nu}) 
(1+ \gamma_5) (\hat {p_K} + \hat {p_{\pi}}) v( p_{\mu} )\;,\\ 
T(\pi^0 \to \gamma \gamma) &=& \frac {\alpha \sqrt 2} { \pi F} 
\epsilon_{\mu \nu \lambda \sigma} k_1^{\mu} e_1^{\nu} k_2^{\alpha} e_2^{\sigma},
\end{eqnarray}
where $F=132$ MeV. It should be noted 
that $T(\pi^0 \to \gamma \gamma)$ is written at $O( p^2)$ level.
In addition, the amplitude differs from that one in [8] by the sign,
since  we used the opposite sign of pseudoscalar octet of mesons. 
From Eq. (17) one can write down the imaginary part of diagram shown
in Fig. 2g:

\begin{eqnarray}
\nonumber 
\mbox {Im} M_3  = \frac {G_F  \alpha} {8 \sqrt 2 \pi^3 F} e \int \frac{d^3 k_{ \pi}} 
{2 \omega_{ \pi}}
\frac{d^3 k_{\mu}} {2 \omega_{\mu}} \delta (k_{\pi}+k_{\mu}-P) 
\frac { \epsilon^{ \rho \sigma \alpha \beta} q_{\alpha} e^*_{\beta} 
k^{\pi}_{\rho} 
 } {k_{\gamma}^2}  \\
\bar u (p_{\nu}) (1+\gamma_5) (\hat p_K + \hat k_{\pi}) (\hat k_{\mu}-m_{\mu})
\gamma_{\sigma} v(p_{\mu})\;. 
\end{eqnarray}
The details of the calculations of integrals entering Eqs. (13), (14), (18) , and 
their dependence on kinematical parameters are given in Appendix~1.

The expression for the amplitude including the  imaginary one-loop contributions
can be written as:
\begin{eqnarray}
M=ie \frac{G_F}{\sqrt{2}}V^{*}_{us}\varepsilon^{*}_{\mu}\Biggl(\tilde f_K
m_{\mu}
\overline u(p_{\nu})(1+\gamma_{5}) \biggl( \frac{p_{K}^{\mu}}{(p_K q)}
-\frac{(p_{\mu})^{\mu}}{(p_{\mu} q)}
\biggr)v(p_{\mu})+ \nonumber\\
\tilde F_n \overline u(p_{\nu})(1+\gamma_{5}) \hat{q} \gamma^{\mu} v(p_{\mu})
-\tilde G^{\mu \nu} l_{\nu}\Biggr),\hspace{1.in}
\end{eqnarray}
where 
\begin{equation}
\tilde G^{\mu \nu}= i \tilde F_v ~\varepsilon^{ \mu \nu \alpha \beta }
q_{\alpha} (p_K)_{\beta} -
\tilde F_a ~ ( g^{\mu \nu} (p_K q)-p_K^{\mu} q^{\nu} ) \;.
\end{equation}
The $\tilde f_K ,\;  \tilde F_v ,\; \tilde F_a$, and 
$\tilde F_n$ formfactors include one-loop contributions 
from diagrams shown in Figs.~2a-2f.  
The choice of the formfactors is determined by the matrix element expansion
into set of gauge-invariant structures.

As long as we are interested in the contributions of 
imaginary parts of one-loop diagrams only (since they lead to a
nonvanishing value of the transverse polarization), 
we neglect the real parts of these diagrams and assume 
that $\mbox{Re} \tilde f_K ,\; \mbox{Re} \tilde F_v ,\; \mbox{Re}
\tilde F_a $ coincide with their tree-level values,
$f_K ,\; F_v ,\;  F_a $, correspondingly, and  
$ \mbox{Re} \tilde F_n= - f_K m_\mu /2(p_\mu q)$.
Explicit expressions for  imaginary parts of the formfactors
are given in Appendix~2.

The muon transverse polarization can be written as
\begin{equation}
P_T=\frac {\rho_T} {\rho_0},
\end{equation}
where
\begin{eqnarray}
\rho_T=2 m_K^3 e^2 G_F^2 |V_{us}|^2 x \sqrt{\lambda y - \lambda^2 - r_{\mu}}
\biggl(
m_{\mu} ~\hbox{Im}(\tilde f_K \tilde F_a^*)(1-\frac{2}{x}+
\frac{y}{\lambda x}) \nonumber\\ 	
+ m_{\mu} ~\hbox{Im}(\tilde f_K \tilde F_v^*)(\frac{y}{\lambda x}-1-2
\frac{r_{\mu}}{\lambda x})+
2 \frac{r_{\mu}}{\lambda x} ~\hbox{Im}(\tilde f_K \tilde F_n^*)(1-\lambda)
\hspace {0.4in}\nonumber\\  
+ m_K^2 x ~\hbox{Im}(\tilde F_n \tilde F_a^*)(\lambda-1)+ m_K^2 x ~\hbox{Im}
(\tilde F_n \tilde F_v^*)(\lambda-1)\biggr)\;.		
\hspace{0.6in}
\end{eqnarray}
It should be noted that Eq. (20) 
disagrees with the expression for $\rho_T$ in [9]. In particular, 
the terms containing Im$F_n$ are missed in the 
$\rho_T$ expression given in [9].
Moreover, calculating the muon transverse polarization we
took into account the diagrams shown in Fig. 2e-g, which have been neglected
in [9], and which give the contribution comparable with
the contribution from other diagrams in Fig. 2. 

\section{Results and discussion}

\noindent
For the numerical calculations we use the following formfactor values
$$ 
f_K=0.16 \mbox{ GeV},\;\; F_v=\frac{0.095}{m_K},\;\; F_a=-\frac{0.043}{m_K}\;.
$$
The $ f_K $ formfactor is determined from experimental data on kaon decays [10],
and $F_v, F_a $ ones are calculated at the one loop-level in the Chiral
Perturbation Theory [11]. It should be noted that our definition for $F_v$
differs by a sign from that in [11]. With this choice of formfactor values the 
decay branching ratio, Br($K^{\pm}\to\mu^{\pm}\nu\gamma$), with the cut on 
photon energy $E_\gamma \geq 20$~MeV, is equal to 
 $=3.3\times 10^{-3}$, which is in good agreement with the PDG data.

The three-dimensional distribution of muon transverse polarization, 
calculated in the one-loop approximation of SM is shown in Fig. 3.
$P_T$, as function of the $x$ and $y$ parameters,
is characterized  by the sum of individual contributions of diagrams in Figs. 2a-f, 
while the contributions from diagrams 2a-d [12] and  2e-f are comparable in absolute 
value, but they are opposite in sign, so that the total $P_T (x,y)$ distribution is
the difference of these group contributions and in absolute value
it is about one order of magnitude less than each individual one of those.

It should be noted that the value of muon transverse polarization
is positive in the whole Dalitz plot region.
Averaged value of transverse polarization can be obtained by integrating the function
$ 2 \rho_T/\Gamma(K^+\to\mu^+\nu\gamma)$
over the physical region, and  with the cut on photon energy $E_\gamma > 20 $~MeV
it is equal to
\begin{equation}
\langle P^{SM}_T \rangle = 5.63 \times 10^{-4}\;. \label{resul1}
\end{equation}
%%%%%%%%%%%%%%%%%%%%%%%%%%%%%%%%%%%%

Let us note that the obtained numerical value of the averaged  transverse polarization
and $P_T(x,y)$ kinematical dependence in Dalitz plot differ from those given
in [9,13]. Note that in [13] only the diagram shown in Fig. 2g was 
calculated and the result for that diagram does not coincide with 
ours. 

As it was calculated in [9], the $P_T$ value varies in the range
of $(-0.1\div 4.0)\cdot 10^{-3}$ for cuts on the muon and photon energies,
$200<E_\mu<254.5$ MeV, $20<E_\gamma<200$ MeV. We have already mentioned above that

1) The authors of [9]  did not take into account  
terms containing the imaginary part of  the $F_n$ formfactor
(contributing to $\rho_T$),
which, in general, are not small being compared with others.

2) The authors of [9] omitted the diagrams, shown in Fig. 2e-f, though, 
as it was mentioned above, their contribution to $P_T$ is comparable with that one
of diagrams in Fig. 2a-2d.

3) The authors of [9] did not take into account the diagram shown in Fig. 2g.

\noindent
All these points lead to serious disagreement between our results and results
obtained in [9]. In particular, our calculations show that the value of
the muon transverse polarization has positive sign in whole Dalitz plot region
and its absolute value varies in the range of $(0.0\div 1.5)\cdot 10^{-3}$, and  
the $P_T$ dependence on the $x, y$ parameters is different from that in [9].

We would like to remark that the muon transverse polarization for the same process 
was calculated in [14], where the contributions from diagrams 2e, 2f and 2g
were taken into account. However, our result differs from the one obtained in [14]:
$P_T$ value has opposite sign in comparison to ours and in numerical calculation 
the author of [14] used constant $f_{\pi}$ instead of $f_K$ in Eq. (1). Since the 
calculation is produced at $O(p^4)$ level, one needs to use $f_K$, as have been
done in our paper. The kinematical structures  for diagrams  Figs. 2a-g in [14]
coincide with ours.

\section*{Acknowledgements}

\noindent
The authors thank Drs. Kiselev V.V. and Likhoded A.K. for fruitful discussion 
and valuable remarks. The authors are also grateful to Bezrukov F., Gorbunov D.
for their remark on sign of formfactor $F_v$ in our previous results
and Rogalyov R. for fruitful discussion. 
This work is in part supported by the Russian Foundation for Basic Research,
grants 99-02-16558 and 00-15-96645, Russian Education Ministry, grant 
RF~E00-33-062 and CRDF MO-011-0. The work of A.A.Likhoded was 
partially funded by a Fapesp grant 2001/06391-4.

\newpage

\normalsize
\vspace*{2cm}
\section*{References}

\vspace*{0.5cm}
\noindent
\begin{description}
\item[1.] S. Weinberg, {\em Phys. Rev. Lett.} {\bf 37} (1976), 651.
\item[2.] A. Likhoded, V. Braguta, A. Chalov, {\em Phys.Rev.} {\bf D65} (2002), 054038.
\item[3.] V.F. Obraztsov and L.G. Landsberg, {\bf hep-ex}/0011033). 
\item[4.] See, for example, {\em Phys. Rev. Lett.} {\bf 83} (1999), 4253;
Yu.G. Kudenko, {\bf hep-ex}/00103007).
\item[5.] J.F. Donoghue and B. Holstein, {\em Phys. Lett.} {\bf B113}(1982),
382;
L. Wolfenstein, {\em Phys. Rev.} {\bf 29} (1984), 2130;
G. Barenboim et al., {\em Phys. Rev.} {\bf 55} (1997), 24213;
M. Kobayashi, T.-T. Lin, Y. Okada, {\em Prog. Theor. Phys.} {\bf 95} (1996),
361; S.S. Gershtein et al., {\em Z. Phys.} {\bf C24} (1984), 305;
R. Garisto, G. Kane, {\em Phys. Rev.} {\bf D44} (1991), 2038.
\item[6.]
G. Belanger, C.Q. Geng, {\em Phys. Rev.} {\bf D44} (1991), 2789.
\item[7.] L.B. Okun and I.B. Khriplovich, {\em Sov. Journ. Nucl. Phys.} {\bf v.6 } (1967), 821.
\item[8.] A.Pich, {\em Rep. Prog. Phys.} {\bf 58} (1995), 563
\item[9.] V.P. Efrosinin, Yu.G. Kudenko, {\em Phys. Atom. Nucl.} {\bf v.62}
(1999), 987.
\item[10.] Review of Particle Physics, {\em Euro. Phys. Journ.} {\bf C15}
(2000). 
\item[11.] J. Bijnens, G. Ecker, J. Gasser, {\em Nucl. Phys.} {\bf B396}
(1993), 81.
\item[12.] A. Likhoded, V. Braguta, A. Chalov,  Preprint IHEP 2000-57, 2000;
\item[13.] G. Hiller and G. Isidori, {\em Phys. Lett.}  {B459} (1999), 295.
\item[14.] R.N. Rogalyov, {\em Phys.Lett.} {\bf B521} (2001), 243

\end{description}

\newpage
\section*{Appendix 1}			  

\noindent
For the the integrals, which contribute to (14) and (15),  we use the following
notations:
\begin{eqnarray}
P&=&p_\mu+q 
\end{eqnarray} 
$$
d \rho =\frac {d^3 k_{\gamma}} {2 \omega_{\gamma}}
\frac {d^3 k_{\mu}} {2 \omega_{\mu}}\delta(k_\gamma+k_{\mu}-P) 
$$
We present below either the explicit expressions for integrals,
or the set of equations, which being solved, give the parameters,
entering the integrals.
\begin{eqnarray}
J_{11}&=&\int d \rho =\frac  {\pi} 2 \frac {P^2-m_{\mu}^2} {P^2}\;,\nonumber \\
J_{12}&=&\int d \rho \frac 1 {(p_K k_{\gamma})}=
\frac {\pi}	{2 I} \ln \biggl( \frac {(P p_K)+I} {(P p_K)-I} \biggr)\;,\nonumber 
\end{eqnarray}
where
$$
I^2=(P p_K)^2-m_K^2 P^2\;,
$$
$$
\int d \rho \frac {k^{\alpha}_{\gamma}} {(p_K k_\gamma)}=a_{11} p^\alpha_K
+b_{11} P^\alpha \;.
$$
The $a_{11}$ and $b_{11}$ parameters are determined by the following
equations:
\begin{eqnarray}
a_{11}&=&-\frac 1 {(P p_K)^2-m_K^2 P^2}
\biggl( P^2 J_{11}- \frac {J_{12}} 2  (P p_K) (P^2-m_{\mu}^2) \biggr)\;,
\nonumber \\
b_{11}&=&\frac 1 {(P p_K)^2-m_K^2 P^2}
\biggl( (P p_K) J_{11}- \frac {J_{12}} 2  m_K^2 (P^2-m_{\mu}^2)
\biggr)\;,\nonumber
\end{eqnarray}
\begin{eqnarray}
\int d \rho k_{\gamma}^{\alpha} &=& a_{12} P^{\alpha}\;, \nonumber \\
\int d \rho k_{\gamma}^{\alpha} k_{\gamma}^{\beta}&=&a_{13} g^{\alpha \beta}
+b_{13} P^{\alpha} P^{\beta}\;, \nonumber 
\end{eqnarray}
where
\begin{eqnarray}
a_{12}&=&\frac {(P^2-m_{\mu}^2)} {2 P^2} J_{11}\;, \nonumber \\
a_{13}&=&-\frac 1 {12} \frac {(P^2-m_{\mu}^2)^2} {P^2} J_{11}\;, \nonumber \\
b_{13}&=&\frac 1 3 \biggl(\frac {P^2-m_{\mu}^2} {P^2} \biggr)^2 J_{11}\;.
\nonumber 
\end{eqnarray}
\begin{eqnarray}
J_1&=&\int d \rho \frac 1 {(p_K k_\gamma)((p_\mu-k_\gamma)^2-m_\mu^2)}=
-\frac \pi {2 I_1 (P^2-m_\mu^2) } \ln 
\biggl( \frac {(p_K p_\mu)+I_1} {(p_K p_\mu)-I_1} \biggr)\;, \nonumber \\
J_2&=&\int d \rho \frac 1 {(p_\mu-k_\gamma)^2-m_\mu^2}=
-\frac \pi {4 I_2} \ln \biggl( \frac {(P p_\mu)+I_2} {(P p_\mu)-I_2} \biggr)
\;, \nonumber 
\end{eqnarray}
where
\begin{eqnarray}
I_1^2&=&(p_K p_\mu)^2 -m_\mu^2 m_K^2\;, \nonumber\\
I_2^2&=&(P p_\mu)^2-m_\mu^2 P^2\;. \nonumber
\end{eqnarray}
\begin{eqnarray}
\int d \rho \frac {k_\gamma^\alpha} {(p_\mu-k_\gamma)^2-m_\mu^2}&=&
a_1 P^\alpha + b_1 p_\mu^\alpha\;, \nonumber \\
a_1&=&-\frac {m_\mu^2 (P^2-m_\mu^2) J_2+(P p_\mu)J_{11}} {2 ((P
p_\mu)^2-m_\mu^2 P^2)} 
\;, \nonumber\\
b_1&=&\frac {(P p_\mu)(P^2-m_\mu^2) J_2+P^2 J_{11}} {2 ((P p_\mu)^2-m_\mu^2
P^2)}
\;, \nonumber
\end{eqnarray}
The integrals below are determined by the parameters, which can be obtained by solving
the sets of equations.
$$
\int d \rho \frac {k_\gamma^\alpha} {(p_K
k_\gamma)((p_\mu-k_\gamma)^2-m_\mu^2)}=
a_2 P^\alpha + b_2 p_K^\alpha +c_2 p_\mu^\alpha \;,
$$
$$
\left\{
\begin{array}{ccc}
a_2 (P p_K)+ b_2 m_K^2+c_2 (p_K p_\mu)=J_2 \hfill \\
a_2 (P p_\mu)+b_2 (p_K p_\mu)+c_2 m_\mu^2=-\frac 1 2 J_{12} \hfill \\
a_2 P^2+b_2 (P p_K)+c_2 (P p_\mu)=(p_\mu q) J_1 \hfill 
\end{array}
\right.
$$
\begin{eqnarray}
\int d \rho \frac {k_\gamma^\alpha k_\gamma^\beta} 
{(p_K k_\gamma)((p_\mu-k_\gamma)^2-m_\mu^2)}&=&
a_3 g^{\alpha \beta}+b_3 (P^\alpha p_K^\beta+P^\beta p_K^\alpha)+
c_3 (P^\alpha p_\mu^\beta+P^\beta p_\mu^\alpha) \nonumber \\
&+& d_3 (p_K^\alpha p_\mu^\beta+p_K^\beta p_\mu^\alpha)+
e_3 p_\mu^\alpha p_\mu^\beta \nonumber \\
&+&f_3 P^\alpha P^\beta +g_3 p_K^\alpha p_K^\beta \;,\nonumber
\end{eqnarray}
\small
$$
\left\{
\begin{array}{cccccccc}
4 a_3+2 b_3 (P p_K)+2 c_3 (P p_\mu)+2 d_3 (p_K p_\mu)+g_3 m_K^2 + e_3
m_\mu^2+f_3 P^2=0 \hfill \\
c_3 (p_K p_\mu) + b_3 m_K^2 + f_3 (P p_K)-a_1=0 \hfill \\
c_3 (P p_K)+d_3 m_K^2+e_3 (p_K p_\mu)-b_1=0 \hfill \\
a_3 + b_3 (P p_K)+d_3 (p_K p_\mu)+g_3 m_K^2=0 \hfill \\
b_3 (p_K p_\mu)+c_3 m_\mu^2+f_3 (P p_\mu)=- \frac 1 2 b_{11} \hfill \\
b_3 (P p_\mu)+d_3 m_\mu^2+g_3 (p_K p_\mu)=-\frac 1 2 a_{11}	 \hfill \\
a_3 P^2+2 b_3 P^2 (P p_K)+2 c_3 P^2 (P p_\mu)+2 d_3 (P p_\mu) (P p_K)+ 
e_3 (P p_\mu)^2+f_3 (P^2)^2+g_3 (P p_K)^2=(p_\mu q)^2J_1 \hfill \\
\end{array}
\right.
$$
\normalsize
$$
\int d \rho \frac {k_\gamma^\alpha k_\gamma^\beta}
{(p_\mu-k_\gamma)^2-m_\mu^2}=
a_4 g_{\alpha \beta}+b_4 (P^\alpha p_\mu^\beta+P^\beta p_\mu^\alpha)+
c_4 P^\alpha P^\beta +d_4 p_\mu^\alpha p_\mu^\beta \;,
$$
$$
\left\{
\begin{array}{cccc}
a_4+d_4 m_\mu^2+b_4 (P p_\mu)=0 \hfill \\
b_4 m_\mu^2+c_4 (P p_\mu)=-\frac 1 2 a_{12} \hfill \\
4 a_4+2 b_4 (P p_\mu)+c_4 P^2+d_4 m_\mu^2=0 \hfill \\
a_4 P^2+2 b_4 P^2 (P p_\mu)+c_4 (P^2)^2+d_4 (P p_\mu)^2==\frac
{(P^2-m_\mu^2)^2} 4 J_2 
\end{array}
\right.
$$

\bigskip

\begin{eqnarray}
\int d \rho \frac {k_\gamma^\alpha k_\gamma^\beta k_\gamma^\delta}
{(p_\mu-k_\gamma)^2-m_\mu^2}&=&a_5 (g^{\alpha \beta} p_\mu^\delta+
g^{ \delta \alpha} p_\mu^\beta+g^{\beta \delta} p_\mu^\alpha)+
b_5 (g^{\alpha \beta} P^\delta +
g^{ \delta \alpha} P^\beta+g^{\beta \delta} P^\alpha) \nonumber \\
&+& c_5 p_\mu^\alpha  p_\mu^\beta  p_\mu^\delta  
+ d_5 P^\alpha P^\beta P^\delta+
e_5 (P^\alpha p_\mu^\beta p_\mu^\delta+P^\delta p_\mu^\alpha p_\mu^\beta+
P^\beta p_\mu^\delta p_\mu^\alpha)\nonumber \\
&+& f_5 (P^\alpha P^\beta p_\mu^\delta+P^\delta P^\alpha p_\mu^\beta+
P^\beta P^\delta p_\mu^\alpha) \;,\nonumber
\end{eqnarray}
\small
$$
\left\{
\begin{array}{cccccc} 
2 a_5+c_5 m_\mu^2+e_5 (P p_\mu)=0 \hfill \\
a_5 m_\mu^2+b_5 (P p_\mu)=-\frac {1}{2}a_{13} \hfill \\
b_5 + e_5 m_\mu^2+f_5 (P p_\mu)=0 \hfill \\
d_5 (P p_\mu)+f_5 m_\mu^2=-\frac 1 2 b_{13} \hfill \\
6 a_5+c_5 m_\mu^2+2 e_5 (P p_\mu)+f_5 P^2=0 \hfill \\
3 a_5   P^2 (P p_\mu)+3 b_5 ( P^2)^2 +c_5 (P p_\mu)^3+
d_5 (P^2)^3+3 e_5 P^2  (P p_\mu)^2+
3 f_5 (P^2)^2 (P p_\mu)=\frac {(P^2-m_\mu^2)^3} 8 J_2
\end{array}
\right.
$$
For the rest of integrals the following notations are used:

\begin{eqnarray}
\nonumber
Pk_{\pi} &=& \frac 1 2 ( P^2 + m_{\pi}^2 - m_{\mu}^2) \\ \nonumber
\end{eqnarray} 
$$
d \rho =\frac {d^3 k_{\pi}} {2 \omega_{\pi}}
\frac {d^3 k_{\mu}} {2 \omega_{\mu}}\delta(k_\pi+k_{\mu}-P) 
$$

In terms of this notations the integrals can be rewritten as follows 

$$
J_3 = \int \frac {d \rho} {k_{\gamma}^2}= - 
\frac {\pi} {4 P q} \mbox {Ln} | \frac { 2 (Pq) Pk_{\pi} + 2 (Pq) 
\sqrt {Pk_{\pi}^2 -m_{\pi}^2 P^2}-m_{\pi}^2 P^2} { 2 (Pq) Pk_{\pi} - 2 (Pq) 
\sqrt {Pk_{\pi}^2 -m_{\pi}^2 P^2}-m_{\pi}^2 P^2}|
$$
$$
J_4=  \int {d \rho} = \frac  {\pi} {P^2} \sqrt {(P k_{\pi})^2-m_{\pi}^2 P^2}
$$

\normalsize
\newpage
\section*{Appendix 2}

\noindent
Here we present the expressions for imaginary parts of form-factors
as the  functions of parameters, calculated in Appendix 1.

\begin{eqnarray}
{\hbox{Im}\tilde{f}_K}&=&\frac{\alpha}{2\pi} {f_K}\
\big(-4\ {a_3}\
{(p_Kq)}+4\ {a_2}\ {{{m_{\mu}}}^2}\
{(p_Kq)}-2\ {b_3}\ {{{m_{\mu}}}^2}\
{(p_Kq)}+4\ {c_2}\ {{{m_{\mu}}}^2}\
{(p_Kq)}-  \nonumber \\
&&  4\ {c_3}\ {{{m_{\mu}}}^2}\
{(p_Kq)}-2\ {d_3}\ {{{m_{\mu}}}^2}\
{(p_Kq)}-2\ {e_3}\ {{{m_{\mu}}}^2}\
{(p_Kq)}-2\ {f_3}\ {{{m_{\mu}}}^2}\
{(p_Kq)}+   \nonumber \\
 && 4\ {a_2}\ {(p_Kq)}\ {(p_{\mu}q)}-4\
{b_3}\ {(p_Kq)}\ {(p_{\mu}q)}-4\ {c_3}\
{(p_Kq)}\ {(p_{\mu}q)}-4\ {f_3}\
{(p_Kq)}\ {(p_{\mu}q)}\big)+   \nonumber \\
&&\frac{\alpha}{2\pi} {F_a}\ \big(8\ {a_4}\
{(p_Kq)}-8\
{a_5}\ {(p_Kq)}-8\ {b_5}\ {(p_Kq)}+8\
{b_4}\ {{{m_{\mu}}}^2}\ {(p_Kq)}+   \nonumber \\
&& 4\ {c_4}\ {{{m_{\mu}}}^2}\
{(p_Kq)}-2\ {c_5}\ {{{m_{\mu}}}^2}\
{(p_Kq)}+4\ {d_4}\ {{{m_{\mu}}}^2}\
{(p_Kq)}-2\ {d_5}\ {{{m_{\mu}}}^2}\
{(p_Kq)}-   \nonumber \\
&& 6\ {e_5}\ {{{m_{\mu}}}^2}\
{(p_Kq)}-6\ {f_5}\ {{{m_{\mu}}}^2}\
{(p_Kq)}+12\ {b_4}\ {(p_Kq)}\
{(p_{\mu}q)}+8\ {c_4}\ {(p_Kq)}\
{(p_{\mu}q)}+   \nonumber \\
&& 4\ {d_4}\ {(p_Kq)}\ {(p_{\mu}q)}-4\
{d_5}\ {(p_Kq)}\ {(p_{\mu}q)}-4\ {e_5}\
{(p_Kq)}\ {(p_{\mu}q)}-8\ {f_5}\
{(p_Kq)}\ {(p_{\mu}q)}\big)+   \nonumber \\
&& \frac{\alpha}{2\pi} {F_v}\ \big(8\ {a_4}\
{(p_Kq)}-8\
{a_5}\ {(p_Kq)}-8\ {b_5}\ {(p_Kq)}+8\
{b_4}\ {{{m_{\mu}}}^2}\ {(p_Kq)}+   \nonumber \\
&& 4\ {c_4}\ {{{m_{\mu}}}^2}\
{(p_Kq)}-2\ {c_5}\ {{{m_{\mu}}}^2}\
{(p_Kq)}+4\ {d_4}\ {{{m_{\mu}}}^2}\
{(p_Kq)}-2\ {d_5}\ {{{m_{\mu}}}^2}\
{(p_Kq)}-   \nonumber \\
&& 6\ {e_5}\ {{{m_{\mu}}}^2}\
{(p_Kq)}-6\ {f_5}\ {{{m_{\mu}}}^2}\
{(p_Kq)}+12\ {b_4}\ {(p_Kq)}\
{(p_{\mu}q)}+8\ {c_4}\ {(p_Kq)}\
{(p_{\mu}q)}+   \nonumber \\
&& 4\ {d_4}\ {(p_Kq)}\ {(p_{\mu}q)}-4\
{d_5}\ {(p_Kq)}\ {(p_{\mu}q)}-4\ {e_5}\
{(p_Kq)}\ {(p_{\mu}q)}-8\ {f_5}\
{(p_Kq)}\ {(p_{\mu}q)}\big) \nonumber 
\end{eqnarray}

\begin{eqnarray}
{\hbox{Im}\tilde{F_a}}&=&\frac{\alpha}{2\pi}{f_K}\
\Big({a_2}\
{{{m_{\mu}}}^2}+2\ {c_2}\
{{{m_{\mu}}}^2}-{c_3}\ {{{m_{\mu}}}^2}-2\
{d_3}\ {{{m_{\mu}}}^2}-{e_3}\
{{{m_{\mu}}}^2}- \nonumber \\
&& \frac{{a_1}\
{{{m_{\mu}}}^2}}{{(p_{\mu}q)}}-\frac{{b_1}\
{{{m_{\mu}}}^2}}{{(p_{\mu}q)}}+\frac{2\ {b_4}\
{{{m_{\mu}}}^2}}{{(p_{\mu}q)}}+\frac{{c_4}\
{{{m_{\mu}}}^2}}{{(p_{\mu}q)}}+\frac{{d_4}\
{{{m_{\mu}}}^2}}{{(p_{\mu}q)}}\Big)+ \nonumber  \\
&& \frac{\alpha}{2\pi}{F_v}\ \big(8\ {a_4}-4\
{a_5}-12\
{b_5}-2\ {a_1}\ {{{m_{\mu}}}^2}+4\
{b_4}\ {{{m_{\mu}}}^2}+5\ {c_4}\
{{{m_{\mu}}}^2}-{c_5}\ {{{m_{\mu}}}^2}-  \nonumber \\
&& {d_4}\ {{{m_{\mu}}}^2}-3\ {d_5}\
{{{m_{\mu}}}^2}-5\ {e_5}\ {{{m_{\mu}}}^2}-7\
{f_5}\ {{{m_{\mu}}}^2}+2\ {a_1}\
{(p_Kp_{\mu})}-4\ {b_4}\ {(p_Kp_{\mu})}-  \nonumber \\
&& 4\ {c_4}\ {(p_Kp_{\mu})}+2\ {d_5}\
{(p_Kp_{\mu})}+2\ {e_5}\ {(p_Kp_{\mu})}+4\
{f_5}\ {(p_Kp_{\mu})}+2\ {a_1}\
{(p_Kq)}-  \nonumber \\
&& 2\ {b_4}\ {(p_Kq)}-4\ {c_4}\
{(p_Kq)}+2\ {d_5}\ {(p_Kq)}+2\ {f_5}\
{(p_Kq)}-4\ {a_1}\ {(p_{\mu}q)}+  \nonumber \\
&& 6\ {b_4}\ {(p_{\mu}q)}+10\ {c_4}\
{(p_{\mu}q)}-6\ {d_5}\ {(p_{\mu}q)}-2\
{e_5}\ {(p_{\mu}q)}-8\ {f_5}\
{(p_{\mu}q)}\big)+  \nonumber \\
&& \frac{\alpha}{2\pi}{F_a}\ \big(-6\ {a_4}+2\
{a_5}+{c_4}\ {{{m_{\mu}}}^2}-{d_4}\
{{{m_{\mu}}}^2}-{d_5}\ {{{m_{\mu}}}^2}-  \nonumber \\
&& {e_5}\ {{{m_{\mu}}}^2}-2\ {f_5}\
{{{m_{\mu}}}^2}+2\ {a_1}\ {(p_Kp_{\mu})}-4\
{b_4}\ {(p_Kp_{\mu})}-4\ {c_4}\
{(p_Kp_{\mu})}+  \nonumber \\
&& 2\ {d_5}\ {(p_Kp_{\mu})}+2\ {e_5}\
{(p_Kp_{\mu})}+4\ {f_5}\ {(p_Kp_{\mu})}+2\
{a_1}\ {(p_Kq)}-2\ {b_4}\ {(p_Kq)}- \nonumber  \\
&& 4\ {c_4}\ {(p_Kq)}+2\ {d_5}\
{(p_Kq)}+2\ {f_5}\ {(p_Kq)}+2\ {c_4}\
{(p_{\mu}q)}-2\ {d_5}\ {(p_{\mu}q)}-2\
{f_5}\ {(p_{\mu}q)}\big) \nonumber
\end{eqnarray}

\newpage

\begin{eqnarray}
{\hbox{Im}\tilde{F_n}}&=&\frac{\alpha}{2\pi}{f_K}\ \Big(4\
{a_1}\
{m_{\mu}}+2\ {a_3}\ {m_{\mu}}+2\ {b_1}\
{m_{\mu}}+{b_{11}}\ {m_{\mu}}-2\ {b_4}\
{m_{\mu}}-2\ {c_4}\ {m_{\mu}}-  \nonumber \\
&& {J_{12}}\ {m_{\mu}}-2\ {J_2}\
{m_{\mu}}-{b_2}\ {{{m_K}}^2}\
{m_{\mu}}+{g_3}\ {{{m_K}}^2}\
{m_{\mu}}-2\ {a_2}\ {{{m_{\mu}}}^3}-  \nonumber \\
&& {c_2}\ {{{m_{\mu}}}^3}+{c_3}\
{{{m_{\mu}}}^3}+{f_3}\ {{{m_{\mu}}}^3}-2\
{a_2}\ {m_{\mu}}\ {(p_Kp_{\mu})}-2\
{b_2}\ {m_{\mu}}\ {(p_Kp_{\mu})}+  \nonumber \\
&& 2\ {b_3}\ {m_{\mu}}\
{(p_Kp_{\mu})}-2\ {c_2}\ {m_{\mu}}\
{(p_Kp_{\mu})}+2\ {d_3}\ {m_{\mu}}\
{(p_Kp_{\mu})}+2\ {J_1}\ {m_{\mu}}\
{(p_Kp_{\mu})}+ \nonumber  \\
&& 2\ {b_3}\ {m_{\mu}}\
{(p_Kq)}-\frac{{a_{12}}\
{{{m_{\mu}}}^3}}{{{{(p_{\mu}q)}}^2}}-\frac{{J_{11
}}\
{{{m_{\mu}}}^3}}{{{{(p_{\mu}q)}}^2}}-\frac{{a_{12
}}\ {m_{\mu}}}{{(p_{\mu}q)}}-\frac{2\ {a_4}\
{m_{\mu}}}{{(p_{\mu}q)}}+  \nonumber \\
&& \frac{{J_{11}}\
{m_{\mu}}}{{(p_{\mu}q)}}-\frac{{a_{11}}\
{{{m_K}}^2}\ {m_{\mu}}}{2\ {(p_{\mu}q)}}+\frac{3\
{a_1}\ {{{m_{\mu}}}^3}}{{(p_{\mu}q)}}+\frac{3\
{b_1}\
{{{m_{\mu}}}^3}}{{(p_{\mu}q)}}+\frac{{b_{11}}\
{{{m_{\mu}}}^3}}{2\ {(p_{\mu}q)}}-  \nonumber \\
&& \frac{2\ {J_2}\
{{{m_{\mu}}}^3}}{{(p_{\mu}q)}}-\frac{{b_{11}}\
{m_{\mu}}\
{(p_Kp_{\mu})}}{{(p_{\mu}q)}}+\frac{{J_{12}}\
{m_{\mu}}\
{(p_Kp_{\mu})}}{{(p_{\mu}q)}}-\frac{{b_{11}}\
{m_{\mu}}\ {(p_Kq)}}{{(p_{\mu}q)}}+ \nonumber  \\
&& \frac{{J_{12}}\ {m_{\mu}}\
{(p_Kq)}}{{(p_{\mu}q)}}-2\ {a_2}\
{m_{\mu}}\ {(p_{\mu}q)}+2\ {c_3}\
{m_{\mu}}\ {(p_{\mu}q)}+2\ {f_3}\
{m_{\mu}}\ {(p_{\mu}q)}\Big)+  \nonumber \\
&&\frac{\alpha}{2\pi} {F_v}\ \Big(2\ {a_4}\
{m_{\mu}}-4\
{a_5}\ {m_{\mu}}+2\ {b_{13}}\
{m_{\mu}}-4\ {b_5}\ {m_{\mu}}-  \nonumber \\
&& 2\ {a_1}\ {{{m_{\mu}}}^3}+{c_4}\
{{{m_{\mu}}}^3}-{c_5}\
{{{m_{\mu}}}^3}-{d_4}\
{{{m_{\mu}}}^3}-{d_5}\ {{{m_{\mu}}}^3}-3\
{e_5}\ {{{m_{\mu}}}^3}-  \nonumber \\
&& 3\ {f_5}\ {{{m_{\mu}}}^3}+2\ {a_1}\
{m_{\mu}}\ {(p_Kp_{\mu})}-2\ {c_4}\
{m_{\mu}}\ {(p_Kp_{\mu})}+2\ {d_4}\
{m_{\mu}}\ {(p_Kp_{\mu})}+  \nonumber \\
&& 2\ {d_5}\ {m_{\mu}}\
{(p_Kp_{\mu})}+2\ {e_5}\ {m_{\mu}}\
{(p_Kp_{\mu})}+4\ {f_5}\ {m_{\mu}}\
{(p_Kp_{\mu})}-2\ {c_4}\ {m_{\mu}}\
{(p_Kq)}+  \nonumber \\
&& 2\ {d_5}\ {m_{\mu}}\ {(p_Kq)}+2\
{f_5}\ {m_{\mu}}\ {(p_Kq)}+\frac{3\
{a_{13}}\
{m_{\mu}}}{{(p_{\mu}q)}}+\frac{{b_{13}}\
{{{m_{\mu}}}^3}}{{(p_{\mu}q)}}- \nonumber  \\
&& \frac{{b_{13}}\ {m_{\mu}}\
{(p_Kp_{\mu})}}{{(p_{\mu}q)}}-\frac{{b_{13}}\
{m_{\mu}}\ {(p_Kq)}}{{(p_{\mu}q)}}-2\
{a_1}\ {m_{\mu}}\ {(p_{\mu}q)}+2\
{c_4}\ {m_{\mu}}\ {(p_{\mu}q)}-  \nonumber \\
&& 2\ {d_4}\ {m_{\mu}}\
{(p_{\mu}q)}-2\ {d_5}\ {m_{\mu}}\
{(p_{\mu}q)}-2\ {e_5}\ {m_{\mu}}\
{(p_{\mu}q)}-4\ {f_5}\ {m_{\mu}}\
{(p_{\mu}q)}\Big)+  \nonumber \\
&& \frac{\alpha}{2\pi}{F_a}\ \Big(-6\ {a_4}\
{m_{\mu}}+8\
{a_5}\ {m_{\mu}}-{b_{13}}\ {m_{\mu}}+8\
{b_5}\ {m_{\mu}}-4\ {b_4}\
{{{m_{\mu}}}^3}- \nonumber  \\
&& 2\ {c_4}\ {{{m_{\mu}}}^3}+{c_5}\
{{{m_{\mu}}}^3}-2\ {d_4}\
{{{m_{\mu}}}^3}+{d_5}\ {{{m_{\mu}}}^3}+3\
{e_5}\ {{{m_{\mu}}}^3}+3\ {f_5}\
{{{m_{\mu}}}^3}+  \nonumber \\
&& 2\ {a_1}\ {m_{\mu}}\
{(p_Kp_{\mu})}-2\ {c_4}\ {m_{\mu}}\
{(p_Kp_{\mu})}+2\ {d_4}\ {m_{\mu}}\
{(p_Kp_{\mu})}+2\ {d_5}\ {m_{\mu}}\
{(p_Kp_{\mu})}+ \nonumber  \\
&& 2\ {e_5}\ {m_{\mu}}\
{(p_Kp_{\mu})}+4\ {f_5}\ {m_{\mu}}\
{(p_Kp_{\mu})}-2\ {c_4}\ {m_{\mu}}\
{(p_Kq)}+2\ {d_5}\ {m_{\mu}}\ {(p_Kq)}+
\nonumber \\
&& 2\ {f_5}\ {m_{\mu}}\
{(p_Kq)}-\frac{3\ {a_{13}}\
{m_{\mu}}}{{(p_{\mu}q)}}-\frac{{b_{13}}\
{{{m_{\mu}}}^3}}{2\ {(p_{\mu}q)}}-\frac{{b_{13}}\
{m_{\mu}}\ {(p_Kp_{\mu})}}{{(p_{\mu}q)}}-  \nonumber \\
&& \frac{{b_{13}}\ {m_{\mu}}\
{(p_Kq)}}{{(p_{\mu}q)}}-6\ {b_4}\
{m_{\mu}}\ {(p_{\mu}q)}-4\ {c_4}\
{m_{\mu}}\ {(p_{\mu}q)}-2\ {d_4}\
{m_{\mu}}\ {(p_{\mu}q)}+  \nonumber \\
&& 2\ {d_5}\ {m_{\mu}}\
{(p_{\mu}q)}+2\ {e_5}\ {m_{\mu}}\
{(p_{\mu}q)}+4\ {f_5}\ {m_{\mu}}\
{(p_{\mu}q)}\Big)\nonumber 
\end{eqnarray}

\newpage

\begin{eqnarray}
{\hbox{Im}\tilde{F_v}}&=&\frac{\alpha}{2\pi}{f_K}\
\Big({a_2}\
{{{m_{\mu}}}^2}+{c_3}\
{{{m_{\mu}}}^2}+{e_3}\ {{{m_{\mu}}}^2}+ \nonumber  \\
&& \frac{{a_1}\
{{{m_{\mu}}}^2}}{{(p_{\mu}q)}}+\frac{{b_1}\
{{{m_{\mu}}}^2}}{{(p_{\mu}q)}}-\frac{2\ {b_4}\
{{{m_{\mu}}}^2}}{{(p_{\mu}q)}}-\frac{{c_4}\
{{{m_{\mu}}}^2}}{{(p_{\mu}q)}}-\frac{{d_4}\
{{{m_{\mu}}}^2}}{{(p_{\mu}q)}}\Big)+ \nonumber  \\
&& \frac{\alpha}{2\pi}{F_a}\ \big(6\ {a_4}-2\
{a_5}-8\
{b_5}+{c_4}\ {{{m_{\mu}}}^2}-{d_4}\
{{{m_{\mu}}}^2}-{d_5}\
{{{m_{\mu}}}^2}-{e_5}\ {{{m_{\mu}}}^2}-  \nonumber \\
&& 2\ {f_5}\ {{{m_{\mu}}}^2}-2\ {a_1}\
{(p_Kp_{\mu})}+4\ {b_4}\ {(p_Kp_{\mu})}+4\
{c_4}\ {(p_Kp_{\mu})}-2\ {d_5}\
{(p_Kp_{\mu})}- \nonumber  \\
&& 2\ {e_5}\ {(p_Kp_{\mu})}-4\ {f_5}\
{(p_Kp_{\mu})}-2\ {a_1}\ {(p_Kq)}+2\
{b_4}\ {(p_Kq)}+4\ {c_4}\ {(p_Kq)}-  \nonumber \\
&& 2\ {d_5}\ {(p_Kq)}-2\ {f_5}\
{(p_Kq)}+2\ {c_4}\ {(p_{\mu}q)}-2\
{d_5}\ {(p_{\mu}q)}-2\ {f_5}\
{(p_{\mu}q)}\big)+  \nonumber \\
&& \frac{\alpha}{2\pi}{F_v}\ \big(-8\ {a_4}+4\
{a_5}+4\
{b_5}+2\ {a_1}\ {{{m_{\mu}}}^2}-4\
{b_4}\ {{{m_{\mu}}}^2}-3\ {c_4}\
{{{m_{\mu}}}^2}+{c_5}\ {{{m_{\mu}}}^2}-  \nonumber \\
&& {d_4}\ {{{m_{\mu}}}^2}+{d_5}\
{{{m_{\mu}}}^2}+3\ {e_5}\ {{{m_{\mu}}}^2}+3\
{f_5}\ {{{m_{\mu}}}^2}-2\ {a_1}\
{(p_Kp_{\mu})}+4\ {b_4}\ {(p_Kp_{\mu})}+ \nonumber  \\
&& 4\ {c_4}\ {(p_Kp_{\mu})}-2\ {d_5}\
{(p_Kp_{\mu})}-2\ {e_5}\ {(p_Kp_{\mu})}-4\
{f_5}\ {(p_Kp_{\mu})}-2\ {a_1}\
{(p_Kq)}+ \nonumber  \\
&& 2\ {b_4}\ {(p_Kq)}+4\ {c_4}\
{(p_Kq)}-2\ {d_5}\ {(p_Kq)}-2\ {f_5}\
{(p_Kq)}+4\ {a_1}\ {(p_{\mu}q)}-  \nonumber \\
&& 6\ {b_4}\ {(p_{\mu}q)}-6\ {c_4}\
{(p_{\mu}q)}+2\ {d_5}\ {(p_{\mu}q)}+2\
{e_5}\ {(p_{\mu}q)}+4\ {f_5}\
{(p_{\mu}q)}\big) \nonumber 
\end{eqnarray}

The contribution to imaginary parts coming from diagram shown in 
Fig. 2g may be written as follows:

$$
\mbox {Im} \tilde f_K=0
$$

\begin{eqnarray}
\nonumber
\mbox {Im} \tilde F_v =- \frac {\alpha} {8 \pi^3 F} 
\left (
\frac 
{3 J_4 m_{\mu}^2} {4 P^2} - \frac {J_4 m_{\mu}^4 m_{\pi}^2} {8 (p_{\mu}q)^2 P^2} +
\frac {J_3 m_{\mu}^4 m_{\pi}^4} {8 (p_{\mu}q)^2 P^2} - \right. \\
\nonumber \left.  - \frac {3 J_4 m_{\mu}^2 m_{\pi}^2} {8 (p_{\mu}q) P^2} + 
\frac {J_3 m_{\mu}^2 m_{\pi}^4} {4 (p_{\mu}q) P^2} + \frac {2 J_4 (p_{\mu}q)} {P^2} 
\right ) \theta ( P^2 - (m_{\mu}+m_{\pi})^2)
\end{eqnarray}
$$
\mbox {Im} \tilde F_a = - \mbox {Im} \tilde F_v
$$

\begin{eqnarray}
\nonumber
\mbox {Im} \tilde f_n = -\frac {\alpha} {8 \pi^3 F} \left (
\frac {-3 J_4 m_{\mu} m_{\pi}^2} {2 P^2} + 
\frac {J_3 m_{\mu} m_{\pi}^4} {P^2} - \frac {J_4 m_{\mu}^5 m_{\pi}^2} {4 (p_{\mu}q)^2 P^2} +
\right.
\\
\nonumber \left. +
 \frac{J_3 m_{\mu}^5 m_{\pi}^4} {4 (p_{\mu}q)^2 P^2} 
- \frac {5 J_4 m_{\mu}^3 m_{\pi}^2} {4 (p_{\mu}q) P^2} + 
\frac {J_3 m_{\mu}^3 m_{\pi}^4} {(p_{\mu}q) P^2} \right )
\theta ( P^2 - (m_{\mu}+m_{\pi})^2)
\end{eqnarray}

\newpage
\section*{Figure captions}

\vspace*{1.5cm}
\noindent
\begin{description}
\item[Fig. 1.] Feynman diagrams for the  $K^{\pm}\to \mu^{\pm}\nu\gamma$ decay
at tree level of SM.
\item[Fig. 2.] Feynman diagrams contributing to
the muon transverse polarization at one-loop level of SM.
\item[Fig. 3.]
The 3D Dalitz plot for the muon transverse polarization
as a function of  $x=2E_\gamma /m_K$ and $y=2E_\mu /m_K $ in the one-loop
approximation of SM.
\item[Fig. 4.] Level lines for the Dalitz plot of the muon transverse
polarization $P_T=f(x,y)$.
\end{description}

\newpage
\setlength{\unitlength}{1mm}
\begin{figure}[ph]
\bf
\begin{picture}(150, 200)

\put(10,160){\epsfxsize=10cm \epsfbox{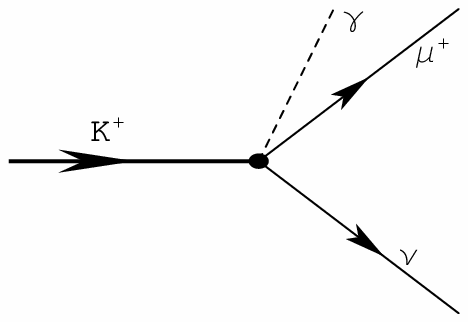}}
\put(28,160){Fig. 1a}

\put(80,160){\epsfxsize=10cm \epsfbox{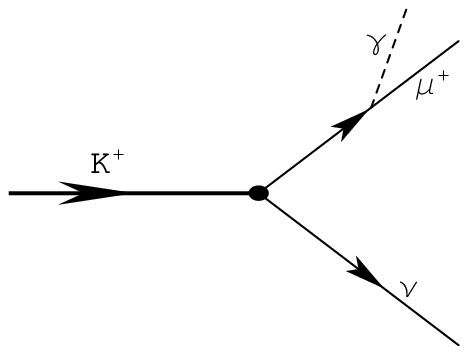}}
\put(98,160){Fig. 1b}

\put(45,60){\epsfxsize=10cm \epsfbox{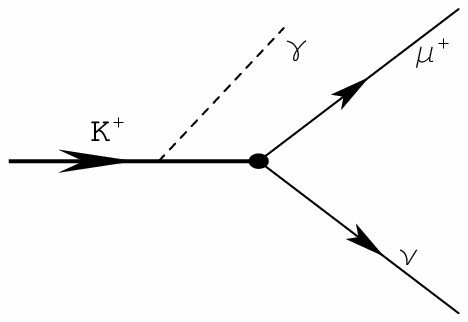}}
\put(64,60){Fig. 1c}

\end{picture}
\end{figure}

\newpage
\setlength{\unitlength}{1mm}
\begin{figure}[ph]
\bf
\begin{picture}(150, 200)

\put(5,170){\epsfxsize=8cm \epsfbox{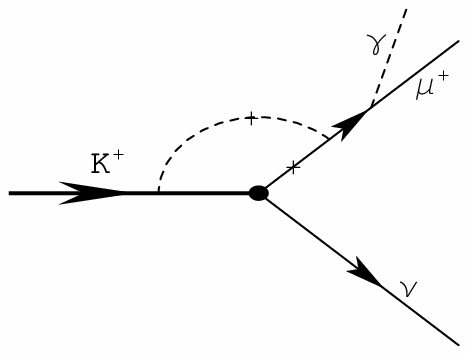}}
\put(23,175){Fig. 2a}

\put(95,170){\epsfxsize=8cm \epsfbox{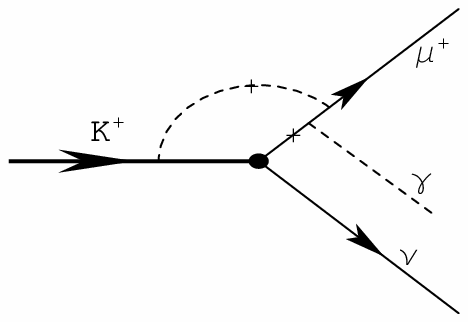}}
\put(123,175){Fig. 2b}

\put(5,110){\epsfxsize=8cm \epsfbox{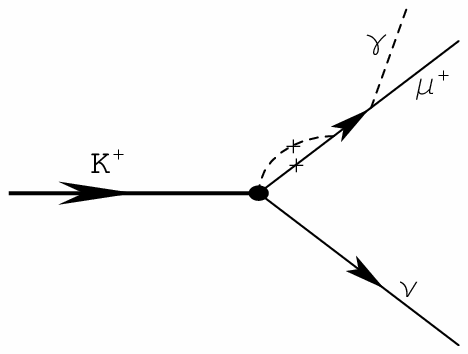}}
\put(23,115){Fig. 2c}

\put(95,110){\epsfxsize=9cm \epsfbox{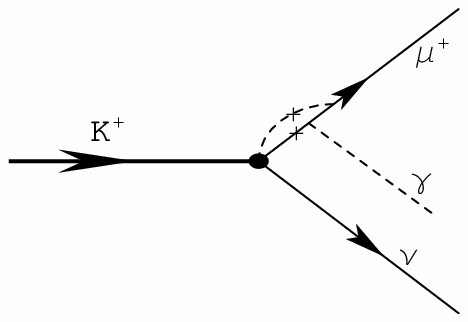}}
\put(123,115){Fig. 2d}

\put(5,50){\epsfxsize=8cm \epsfbox{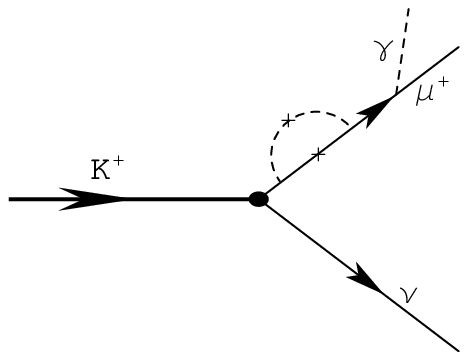}}
\put(23,55){Fig. 2e}

\put(95,50){\epsfxsize=9cm \epsfbox{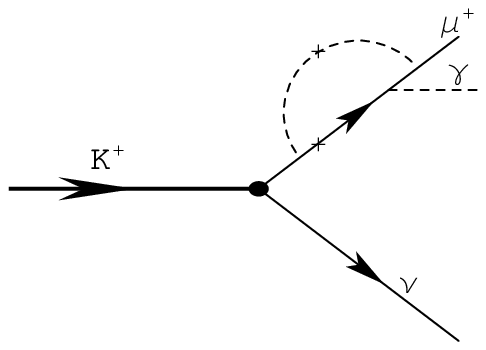}}
\put(123,55){Fig. 2f}

\put(50,0){\epsfxsize=9cm \epsfbox{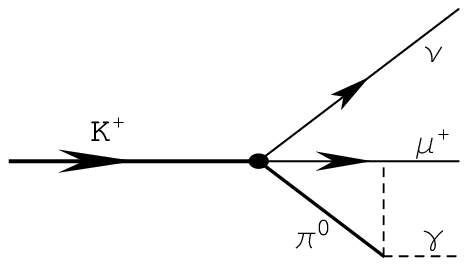}}
\put(73,5){Fig. 2g}

\end{picture}
\end{figure}

\newpage
\setlength{\unitlength}{1mm}
\begin{figure}[ph]
\bf
\begin{picture}(150, 200)
\put(40,130){\epsfxsize=9cm \epsfbox{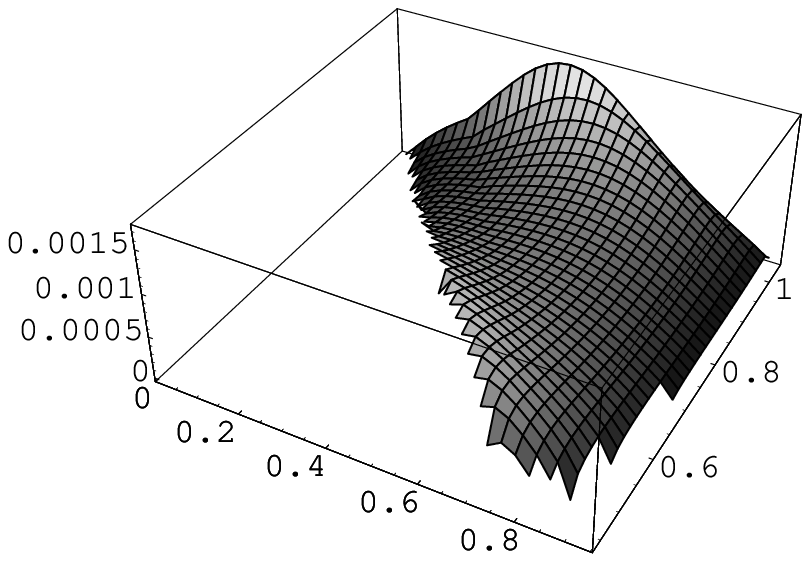}}
\put(80,120){Fig. 3}
\put(75,133){$\bf x$}
\put(125,145){$\bf y$}
\put(35,180){$\bf P_T$}

\put(40,25){\epsfxsize=10cm \epsfbox{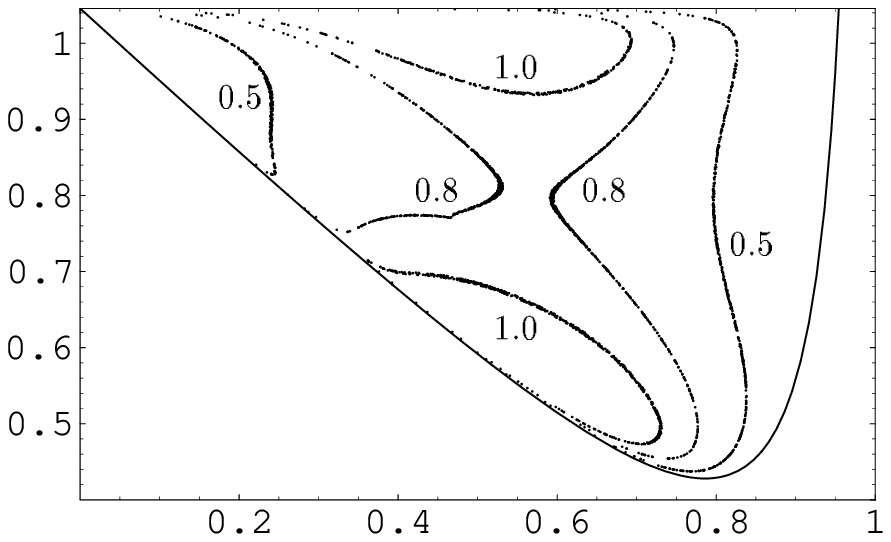}}
\put(90,90){$ P_T\cdot 10^{3}$}
\put(95,21){$\bf x$}
\put(35,60){$\bf y$}

\put(80,10){Fig. 4}

\end{picture}
\end{figure}

\end{document}